**Non-linear swept frequency technique for $CO_2$ measurements using a CW laser system**

Joel F. Campbell
NASA Langley Research Center
Hampton, VA 23281
joel.f.campbell@nasa.gov

## Abstract

A system using a non-linear multi-swept sine wave system is described which employs a multi-channel, multi-swept orthogonal waves, to separate channels and make multiple, simultaneous online/offline $CO_2$ measurements. An analytic expression and systematic method for determining the orthogonal frequencies for the unswept, linear swept and non-linear swept cases is presented. It is shown that one may reduce sidelobes of the autocorrelation function while preserving cross channel orthogonality, for thin cloud rejection.

## Introduction

The U.S. National Research Council's report entitled Earth Science and Applications from Space: National Imperatives for the Next Decade and Beyond identified the need for a near-term space mission of Active Sensing of $CO_2$ Emissions over Nights, Days, and Seasons (ASCENDS) [1]. The primary objective of the ASCENDS mission is to make $CO_2$ column measurements across the troposphere during the day and night over all latitudes and all seasons and in the presence of scattered clouds. These measurements would be used to significantly reduce the uncertainties in global estimates of $CO_2$ sources and sinks; to provide an increased understanding of the connection between climate and $CO_2$ exchange; to improve climate models; and to close the carbon budget for improved forecasting and policy decisions [1].

The ASCENDS mission requires an active sensor that has the needed $CO_2$ measurement precision from space to determine the global variability of $CO_2$ in the troposphere and thereby determine $CO_2$ sources and sinks at the surface. The critical component in the ASCENDS mission is the Laser Absorption Spectrometer (LAS) for $CO_2$ column measurements in the mid to lower troposphere. Potential measurement techniques include pulsed [2-3], pn code [4-5], swept sine wave [6-10], unswept sine wave [11-15], and other similar techniques.

ITT in collaboration with NASA Langley Research Center have been experimenting with linear swept and unswept sine waves as a means of measuring $CO_2$ for many years [9-10, 13-15]. More recently others have done the same using a very similar technique with different processing [8, 12, 16]. In the case of swept frequency, ITT uses a multi-linear swept frequency technique where as many as 1000 sweeps are combined in a single frame. The phase is continuous over the length of the frame where start frequencies are chosen in an ad-hoc fashion to minimize cross correlation cross talk. Using an instrument setup similar to that shown in Figure 1, channels are then separated by cross correlating the return signal with the reference waveforms associated with each channel. One disadvantage to using linear swept frequency is the potential for sidelobe interference when cross correlating the reference signals from multiple targets. The technique presented here minimizes that effect by minimizing the side lobes in a lossless fashion so that signal to noise is preserved. In addition, we find that one may determine the orthogonal frequencies in a systematic way for a more generalized non-linear sweep function.

**Measurement Technique**

For the system described by Figure 1, modulated online and offline lasers are combined, then amplified and transmitted through the beam expander. A portion of this signal is routed to a reference detector and digitized simultaneously with the data received through the science detector. The science detector data comes from ground and other reflections and further processed to discriminate ground and cloud returns. A more elaborate instrument would include sideline wavelengths and modulators as well.

The on-line and off-line modulation signals are

$$\begin{aligned} M_{on} &= 1 + m\cos\left(\phi_1(t)\right), \\ M_{off} &= 1 + m\cos\left(\phi_2(t)\right). \end{aligned} \tag{1}$$

where m is the modulation index and

$$\begin{aligned} \phi_1(t) &= 2\pi\int_0^t f_1(t')\,dt', \\ \phi_2(t) &= 2\pi\int_0^t f_2(t')\,dt', \end{aligned} \tag{2}$$

where $f_1$ and $f_2$ are the online and offline time dependent frequencies. The received optical intensity is given by

$$\begin{aligned} I_{on}(t) &= \frac{K}{r^2}\ P_{on}\exp\left(-2\varepsilon\int_0^r \beta(r')\,dr'\right)\exp(-2\xi)\exp(-2\xi')\left(1 + m\cos\left(\phi_1(t - 2r/c)\right)\right) \\ I_{off}(t) &= \frac{K}{r^2}\ P_{off}\exp\left(-2\varepsilon\int_0^r \beta(r')\,dr'\right)\exp(-2\xi)\left(1 + m\cos\left(\phi_2(t - 2r/c)\right)\right) \end{aligned}, \tag{3}$$

where m is the modulation index, $\xi$ is the total column optical depth for the off-line measurement which is related to the absorption other than $CO_2$, and $\xi'$ is the column optical depth due to the $CO_2$ absorption, $\beta$ is the backscatter coefficient, $\varepsilon$ is the extinction to backscatter ratio, K is a constant, $P_{on}$ is the transmitted on-line power, and $P_{off}$ is the transmitted off-line power, $\phi_1$ is the on-line modulation phase, $\phi_2$ is the off-line modulation phase, and r is the target range. When the optical signals are combined at the detector and converted to an electronic signal and AC coupled we have,

$$S(t) = C_1 \cos\left(\phi_1 \left(t - 2r/c\right)\right) + C_2 \cos\left(\phi_2 \left(t - 2r/c\right)\right), \tag{4}$$

where r is the range to the target and

$$\begin{aligned} C_1 &= \frac{K'}{r^2} P_{on} \exp\left(-2\varepsilon \int_0^z \beta(r')\,dr'\right) \exp(-2\xi) \exp(-2\xi') \\ C_2 &= \frac{K'}{r^2} P_{off} \exp\left(-2\varepsilon \int_0^z \beta(r')\,dr'\right) \exp(-2\xi) \end{aligned}, \tag{5}$$

where K’ is a constant. Solving for $\xi'$ gives

$$\xi' = \frac{1}{2} \ln\left(\frac{C_2 P_{on}}{C_1 P_{off}}\right). \tag{6}$$

Once $C_1$ and $C_2$ are determined the column optical depth for $CO_2$ may be found. In general, this is done by cross correlating the reference waveforms with the return signal. By choosing the reference waveforms carefully such that each channel is orthogonal to every other (perfectly uncorrelated everywhere), one may determine $C_1$, $C_2$, and r directly.

**Multi-swept orthogonal sine waves**

General case

The main advantage of the multi-sweep approach is one may achieve a high level of orthogonality between channels while using simultaneous channels that are very close in frequency. The advantage is that one may make simultaneous on-line/off-line measurements with virtually no crosstalk.

A generalized multi-sweep, multi-channel case may be formulated by first defining the frequency for each channel by

$$f_j(t) = f_{0j} + f_{sweep}\left(t - \tau \ \mathrm{int}(t/\tau)\right), \tag{7}$$

where $j$ is the channel number, $\tau$ is the sweep period, $f_{0i}$ is the start frequency, $f_{sweep}$ is the sweep function, and $\text{int}(t/\tau)$ is the integer part of $t/\tau$. Let

$$\phi_{sweep}(t) = 2\pi \int_0^t f_{sweep}(t')dt' . \quad (8)$$

We now construct a continuous phase function for each channel such that

$$\phi_j(t) = 2\pi f_{0j} t + \phi_{sweep}(\tau)\text{int}(t/\tau) + \phi_{sweep}(t - \tau\ \text{int}(t/\tau)). \quad (9)$$

We digitize these waveforms at a sample rate $f_s$, a sample time of $\Delta t = 1/f_s$ and with a frame size of N points that includes M sweeps such that the total time for M sweeps is $M\tau \equiv N\Delta t$. We define a circular correlation such that

$$\begin{aligned} R(ref, data) &= \frac{1}{N}\sum_{m=0}^{N-1} ref^*(m)\ data(m+n) \\ &= DFT^{-1}\left(DFT^*(ref)DFT(data)\right) \end{aligned}, \quad (10)$$

where ref is the reference waveform, data is the data collected either from the reference or science detector, and DFT is the digital Fourier transform. For the purposes of this paper, ref will take the form of $\exp(i\ \phi_j)$ because in our case the correlation is performed in quadrature. The channel frequencies $f_{0j}$ are chosen to make each channel orthogonal with every other channel such that

$$R\left(\exp(i\ \phi_j), \cos(\phi_k)\right) = 0,\ j \neq k \quad (11)$$

where $i = \sqrt{-1}$. Since each channel is orthogonal, the constants $C_1$ and $C_2$ will then be proportional to the peak of the cross correlation of the return signal with the reference waveform for channels 1 and 2 respectively and the range is determined by the delay in the correlation as in Figure 2 where $r = c\ \delta t/2$ where c is the speed of light and $\delta t$ is the time delay.

We claim the general swept case is closely related to the unswept case (see Appendix) in that the choice of frequencies is similar if we use the average frequency over a single sweep. That is to say one may use average frequency of

$$\langle f_j \rangle = f_{0j} + \langle f_{sweep} \rangle = f_{0j} + \frac{1}{2\pi\tau}\phi_{sweep}(\tau) \quad (12)$$

with the results of the unswept case as a starting point to guess what the orthogonal start frequencies are. Though related, the swept case is somewhat more complicated in the choice of allowable start frequencies. We present as an ansatz a method for choosing start frequencies such that

$$
\begin{aligned}
&f_{01} = \frac{n_1}{2M\tau} - \frac{\phi_{sweep}(\tau)}{2\pi\tau} > 0, \\
&f_{02} = f_{01} + \frac{n_2}{2M\tau}, \\
&\ldots, \\
&f_{0K} = f_{01} + \frac{n_K}{2M\tau},
\end{aligned}
\tag{13}
$$

where K is the number of channels and $n_j$ is an integer. The autocorrelation function for each channel will have exactly M peaks. Generally speaking, if the number of channels is greater than 2, the number of sweeps must be greater than the number of channels (M>K). For the two-channel case we must have at least 2 sweeps. The integer $n_1$ is chosen to make the autocorrelation for the first channel have the same peak height for each sweep. Once this is done, $n_j$ for the other channels is chosen to make each channel orthogonal with every other and the autocorrelation function for each channel has equal peak heights. In a four channel setup, for instance, this can usually be done in a few minutes by plotting the cross correlation between channels and autocorrelation functions.

Linear sweep case

For the linear sweep case we choose

$$f_{sweep}(t) = \Delta f \frac{t}{\tau}, \; 0 \le t < \tau , \tag{14}$$

where $\Delta f$ is the sweep bandwidth. This is a well-known case from radar [17] and will result in a range resolution of

$$\delta r = \frac{c}{2\Delta f} , \tag{15}$$

where c is the speed of light. The unambiguous range is given by

$$r_{\max} = \frac{c \;\tau}{2} . \tag{16}$$

The frequency as a function of time for each channel is

$$f_j(t) = f_{0j} + \Delta f \frac{t - \tau \ \mathrm{int}(t/\tau)}{\tau}, \tag{17}$$

where $\mathrm{int}(t/\tau)$ is the integer part of $t/\tau$. The phase is

$$\phi_j(t) = 2\pi\left[ f_{0j}t + \frac{1}{2}\Delta f \tau \,\mathrm{int}(t/\tau) + \frac{\Delta f}{2\tau}\left(t - \tau \ \mathrm{int}(t/\tau)\right)^2 \right]. \tag{18}$$

The frequency and phase as a function of time are plotted in Figure 3. Figure 4 shows the resulting swept sine wave for the first sweep. In the continuum limit it can be shown the autocorrelation function is

$$\begin{aligned} \left| R\left(\exp(i\ \phi_j), \cos(\phi_j)\right) \right| &\approx \left| \begin{array}{l} \frac{1}{\tau}\int_0^{\tau - t} \exp\left(-i\ \phi_j(t')\right) \exp\left(i\ \phi_j(t'+t)\right) dt' + \\ \frac{1}{\tau}\int_{\tau - t}^{\tau} \exp\left(-i\ \phi_j(t')\right) \exp\left(i\ \phi_j(t'+t)\right) dt' \end{array} \right| \\ &= \left| \frac{\sin\left[\pi\ \Delta f\ t\left(1 - \frac{|t|}{\tau}\right)\right]}{2\pi\ \Delta f\ t\left(1 - \frac{|t|}{\tau}\right)} \right|, -\tau < t < \tau \end{aligned} \tag{19}$$

In the above integration we ignore the lower sideband because that is mostly filtered out in the integration. This is *not* the single sweep pulse compression autocorrelation function normally found in standard radar texts [18, 19]. The difference is that in this case we have back-to-back sweeps so the autocorrelation function is derived from an integration of two sweeps at the same time. This results in an autocorrelation function that has two poles in the denominator instead of one.

As an example we take the special case where we have a sample rate of 2000 kHz, a sweep bandwidth of 500 kHz, a frame size of 4096, and a sweep period of 512/2000 ms which results in M=8 sweeps per frame. We chose 4 channels and the parameters chosen that result in orthogonal sweeps are $n_1$=1450, $n_2$=18, $n_3$=30, and $n_4$=52. This results in start frequencies of approximately 104.004 kHz, 108.398 kHz, 111.328 kHz, and 116.699 kHz. Care must be taken to compute these as accurately as possible or choose parameters that result in start frequencies that are near integers to minimize round off error. Figure 5a and 5b show the autocorrelation function and channel 1-2 cross correlation crosstalk using the above parameters. Cross correlation crosstalk is 0 to within round off error in that case and between all other channels as well.

Non-linear sweep case

The use of non-linear sweeps for use in sidelobe suppression is a well-known technique from radar [20-23]. Although the use of data windows can also reduce sidelobes, it usually results in a loss of signal to noise. Using a well-designed non-linear sweep does not have this limitation. The general idea is to use a sweep that results in a reshaping of the frequency profile of the autocorrelation function in the frequency domain. The desired shape is that of a Gaussian like shape because the Fourier transform of a Gaussian is also a Gaussian, which has no sidelobes. Combining this technique with the results above means we may develop an orthogonal set of such sweeps. One of the simplest techniques we have found for generating a non-linear sweep effective at suppressing sidelobes was that presented by Lesnik[23]. A modification of that technique suitable for this problem results in a sweep function

$$f_{sweep}(t) = \frac{1}{2}\Delta f - \frac{\Delta f\sqrt{1-k^2}}{2}\frac{1-2t/\tau}{\sqrt{1-k^2(1-2t/\tau)^2}} \tag{20}$$

where k is a parameter that determines the strength of the nonlinearity where $0 < k < 1$. By Equation 7 the frequency for the multi-sweep setup is

$$f_j(t) = f_{0j} + \frac{1}{2}\Delta f - \frac{\Delta f\sqrt{1-k^2}}{2}\frac{1-2t'/\tau}{\sqrt{1-k^2(1-2t'/\tau)^2}},\ t' = t - \tau\,\mathrm{int}(t/\tau). \tag{21}$$

The two limits of interest are

$$\lim_{k\to 0} f_j = f_{0j} + \Delta f\frac{t}{\tau},\ \lim_{k\to 1} f_j = f_{0j} + \frac{1}{2}\Delta f\,. \tag{22}$$

As k becomes small it approaches the linear sweep case and as k approaches 1 it approaches unswept case with a frequency equal to the average frequency of the linear swept case.

The phase sweep function is

$$\phi_{sweep}(t) = 2\pi\left[\frac{1}{2}\Delta f t + \frac{\Delta f\tau}{4}\frac{(1-k^2)}{k^2}\left(1-\sqrt{\frac{1-k^2\left(1-\frac{2t}{\tau}\right)^2}{(1-k^2)}}\right)\right]. \tag{23}$$

By Equation 9 the phase is therefore

$$\phi(t) = 2\pi\left[\left(f_{0j} + \frac{1}{2}\Delta f\right)t + \frac{\Delta f \tau}{4}\frac{\left(1-k^2\right)}{k^2}\left(1 - \sqrt{\frac{1-k^2\left(1-\frac{2t'}{\tau}\right)^2}{\left(1-k^2\right)}}\right)\right], \tag{24}$$

$$t' = t - \tau \operatorname{int}\left(t/\tau\right).$$

The exact same start frequencies and $n_j$ may be used for this particular non-linear sweep function. Part of the reason for this is the average frequency over the sweep period is exactly the same as for the linear case. Figure 6a and Figure 6b are plots of $f_{sweep}$ and $\phi_{sweep}$ for various choices of k ranging 0 (linear) to 0.99 (strongly non-linear).

Figure 7 is a comparison between the FFT power spectrum of $\cos\left(\phi_j\right)$ for k=.001 and k=0.99. This plot shows the change from a relatively flat top power spectrum to that of a Gaussian shape. We have also verified the area under the power curve is conserved between both cases which means the transformation is lossless unlike the case of a data window. That means the SNR is also conserved, unlike the case of using a data window. Figure 8 shows that for a sweep length 512 and k=0.91 the highest sidelobe is about -34dB. Figures 9a shows this can be improved by choosing k=0.999 with a sweep length of 4096, which gives a highest sidelobe level of about -55dB. By choosing k=0.99999 with a sweep length of 2 the highest sidelobes drops to below -80dB. The downside is there is a loss of resolution, improved by increasing the sweep bandwidth and increasing the sample rate. In the end we may achieve any reasonable sidelobe level and resolution we wish provided we have the necessary hardware.

Figure 10a and 10b show the autocorrelation function channel 1 and channel 1-2 cross correlation crosstalk for k=0.91. The results here are just as impressive as they are in the linear case. The amazing thing is this was accomplished using the exact same start frequencies as for the linear case. The results are just as impressive for the other channels as well.

**Discussion**

We have presented a technique that is useful in reducing side lobes for the multi-swept frequency, orthogonal multi-channel technique. This technique is flexible in reducing side lobes to within the limits of the hardware used. With a high enough sample rate and sweep bandwidth, one could potentially reduce the side lobes to any level required to a specified resolution. This should be very helpful in thin cloud rejection due to the reduction in sidelobe interference. In the near future we plan to test some of these ideas in hardware. We have also presented an analytic technique for choosing the orthogonal start frequencies for both the linear and non-linear sweep case.

**Acknowledgements**

We thank Jeremy Dobler, Bing Lin, Michael Vanek, Wallace Harrison and Amin Nehrir for their helpful discussions and information.

## Appendix

### Unswept sine wave case

The unswept sine wave modulation case has been described elsewhere [11-15]. It has been used for many years as a technique for measuring $CO_2$. The major disadvantage with this technique is it typically requires a separate altimeter. In situations where thin clouds are present one cannot discriminate between the ground and a thin cloud, which makes the measurement problematic.

Quadrature demodulation [11,13,24] is used to amplitude demodulate the signal in a way makes the measurement insensitive to changes in phase. A digital implementation of this is shown in Figure A1.

The incoming signal from the detector will have the form

$$\begin{aligned} S(t) &= C_1 \cos(2\pi f_1 t + \phi_1) + C_2 \cos(2\pi f_2 t + \phi_2) \\ &= A_1 \sin(2\pi f_1 t) + B_1 \cos(2\pi f_1 t) + A_2 \sin(2\pi f_2 t) + B_2 \cos(2\pi f_2 t) \end{aligned}, \quad \text{(A1)}$$

where

$$\begin{aligned} A_i &= C_i \cos(\phi_i), \\ B_i &= -C_i \sin(\phi_i). \end{aligned} \quad \text{(A2)}$$

First the signal is sampled discretely with a sample time $\Delta t$ such that $S(t) \to S(n\Delta t)$, where n is the sample number. We then we demodulate the signal as shown in Figure A1. For each leg of the quadrature demodulator we find

$$\begin{aligned} x_i &= \frac{1}{N} \sum_{n=0}^{N-1} \sin(2\pi f_i \tau n) S(n\Delta t), \\ y_i &= \frac{1}{N} \sum_{n=0}^{N-1} \cos(2\pi f_i \tau n) S(n\Delta t). \end{aligned} \quad \text{(A3)}$$

These sums can be done analytically using the following relations,

$$
\begin{aligned}
&\frac{1}{N}\sum_{n=0}^{N-1}\sin\left(2\pi f_i n\Delta t\right)\sin\left(2\pi f_j n\Delta t\right) \\
&\quad=\frac{1}{4N}\frac{\sin\left(\left(2N-1\right)\pi\left(f_i-f_j\right)\Delta t\right)}{\sin\left(\pi\left(f_i-f_j\right)\Delta t\right)}-\frac{1}{4N}\frac{\sin\left(\left(2N-1\right)\pi\left(f_i+f_j\right)\Delta t\right)}{\sin\left(\pi\left(f_i+f_j\right)\Delta t\right)}, i\neq j \\
&\quad=\frac{1}{4N}\left(2N-1\right)-\frac{1}{4N}\frac{\sin\left(2\left(2N-1\right)\pi f_i\Delta t\right)}{\sin\left(2\pi f_i\Delta t\right)}, i=j
\end{aligned}
\tag{A4}
$$

$$
\begin{aligned}
&\frac{1}{N}\sum_{n=0}^{N-1}\cos\left(2\pi f_i n\Delta t\right)\cos\left(2\pi f_j n\Delta t\right) \\
&\quad=\frac{1}{2N}+\frac{1}{4N}\frac{\sin\left(\left(2N-1\right)\pi\left(f_i-f_j\right)\Delta t\right)}{\sin\left(\pi\left(f_i-f_j\right)\Delta t\right)}+\frac{1}{4N}\frac{\sin\left(\left(2N-1\right)\pi\left(f_i+f_j\right)\Delta t\right)}{\sin\left(\pi\left(f_i+f_j\right)\Delta t\right)}, i\neq j \\
&\quad=\frac{1}{4N}\left(2N+1\right)+\frac{1}{4N}\frac{\sin\left(2\left(2N-1\right)\pi f_i\Delta t\right)}{\sin\left(2\pi f_i\Delta t\right)}, i=j
\end{aligned}
\tag{A5}
$$

$$
\begin{aligned}
&\frac{1}{N}\sum_{n=0}^{N-1}\cos\left(2\pi f_i n\Delta t\right)\sin\left(2\pi f_j n\Delta t\right) \\
&\quad=-\frac{1}{2N}\frac{\sin\left(N\pi\left(f_i-f_j\right)\Delta t\right)\sin\left(\left(N-1\right)\pi\left(f_i-f_j\right)\Delta t\right)}{\sin\left(\pi\left(f_1-f_2\right)\Delta t\right)} \\
&\quad+\frac{1}{2N}\frac{\sin\left(N\pi\left(f_i+f_j\right)\Delta t\right)\sin\left(\left(N-1\right)\pi\left(f_i+f_j\right)\Delta t\right)}{\sin\left(\pi\left(f_i+f_j\right)\Delta t\right)}, i\neq j \\
&\quad=\frac{1}{2N}\frac{\sin\left(2N\pi f_i\Delta t\right)\sin\left(2\left(N-1\right)\pi f_i\Delta t\right)}{\sin\left(2\pi f_i\Delta t\right)}, i=j
\end{aligned}
\tag{A6}
$$

Each one of the sums given by Equations A4-A6, represents the residual error due to the upper and lower sidebands in the lock-in detection process. The simplest of these is Equation 10. By choosing

$$
\begin{aligned}
&\sin\left(N\pi\left(f_1-f_2\right)\Delta t\right)=0 \\
&\sin\left(N\pi\left(f_1+f_2\right)\Delta t\right)=0
\end{aligned},
\tag{A7}
$$

orthogonality is achieved in Equations A4-A6, though this is not obvious at first site. The solution to Equation A7 is satisfied by choosing frequencies such that

$$\begin{aligned} f_1 &= \frac{1}{2N\Delta t}(p-q) \\ f_2 &= \frac{1}{2N\Delta t}(p+q) \end{aligned} \quad . \tag{A8}$$

where p and q are positive integers such that p>q and

$$\frac{1}{N}(p+q) < 1 . \tag{A9}$$

The above equation is a restatement of the Nyquist criteria where $2f_2<1/\Delta t$. Though not obvious, this choice of frequencies also makes Equations A4 and A5 orthogonal by completely filtering out the upper and lower sidebands with no residual. When the frequencies are chosen as described by Equation A8 it can be shown that

$$\begin{aligned} x_i &= \sum_{n=0}^{N-1} \sin(2\pi f_i n\Delta t) S(n\Delta t) = \frac{A_i}{2} \\ y_i &= \sum_{n=0}^{N-1} \cos(2\pi f_i n\Delta t) S(n\Delta t) = \frac{B_i}{2} \end{aligned} \quad , \tag{A10}$$

so that

$$\begin{aligned} \sqrt{x_1^2 + y_1^2} &= \frac{1}{2}\sqrt{A_1^2 + B_1^2} = \frac{1}{2}C_1 \\ \sqrt{x_2^2 + y_2^2} &= \frac{1}{2}\sqrt{A_2^2 + B_2^2} = \frac{1}{2}C_2 \end{aligned} \quad . \tag{A11}$$

This represents the signal strength of the off-line and on-line signals respectively and can be used to find the column optical depth for $CO_2$ as shown in Equation 4. In the event more channels are needed, candidate frequencies may be chosen from.

**Figure captions**

Figure 1. Simplified instrument block diagram.

Figure 2. Result of cross correlation of the return signal to obtain range and amplitude.

Figure 3. Frequency (a) and phase (b) as a function of time for linear sweep case.

Figure 4. Single linear sweep sine wave.

Figure 5. Channel 1 autocorrelation function (a) and channel 1-2 cross correlation cross talk (b).

Figure 6. Single sweep nonlinear sweep (a) and phase (b) for different values of k.

Figure 7. Comparison between linear and strongly non-linear case shows that as k approaches 1 the frequency profile approaches a Gaussian shape in a way that the area under the power curve is preserved.

Figure 8. Comparison between linear sweep and non-linear sweep for 512 sweep length and k=0.91 shows highest sidelobe in non-linear case is about -34 dB compared to -13 dB in the linear case.

Figure 9. Highest sidelobe is -55 dB for k=0.999 and 4096 points per sweep (a). Highest sidelobe is below -80 dB for k=.99999 and $2^{17}$ points per sweep (b).

Figure 10. Channel 1 nonlinear autocorrelation function for k=0.91 and a sweep length of 512 shows reduced sidelobes (a) and channel 1-2 cross correlation cross talk is as good as the linear case (b).

Figure 1A. Quadrature/lock-in demodulation used for signal detection of unswept sine wave modulation.

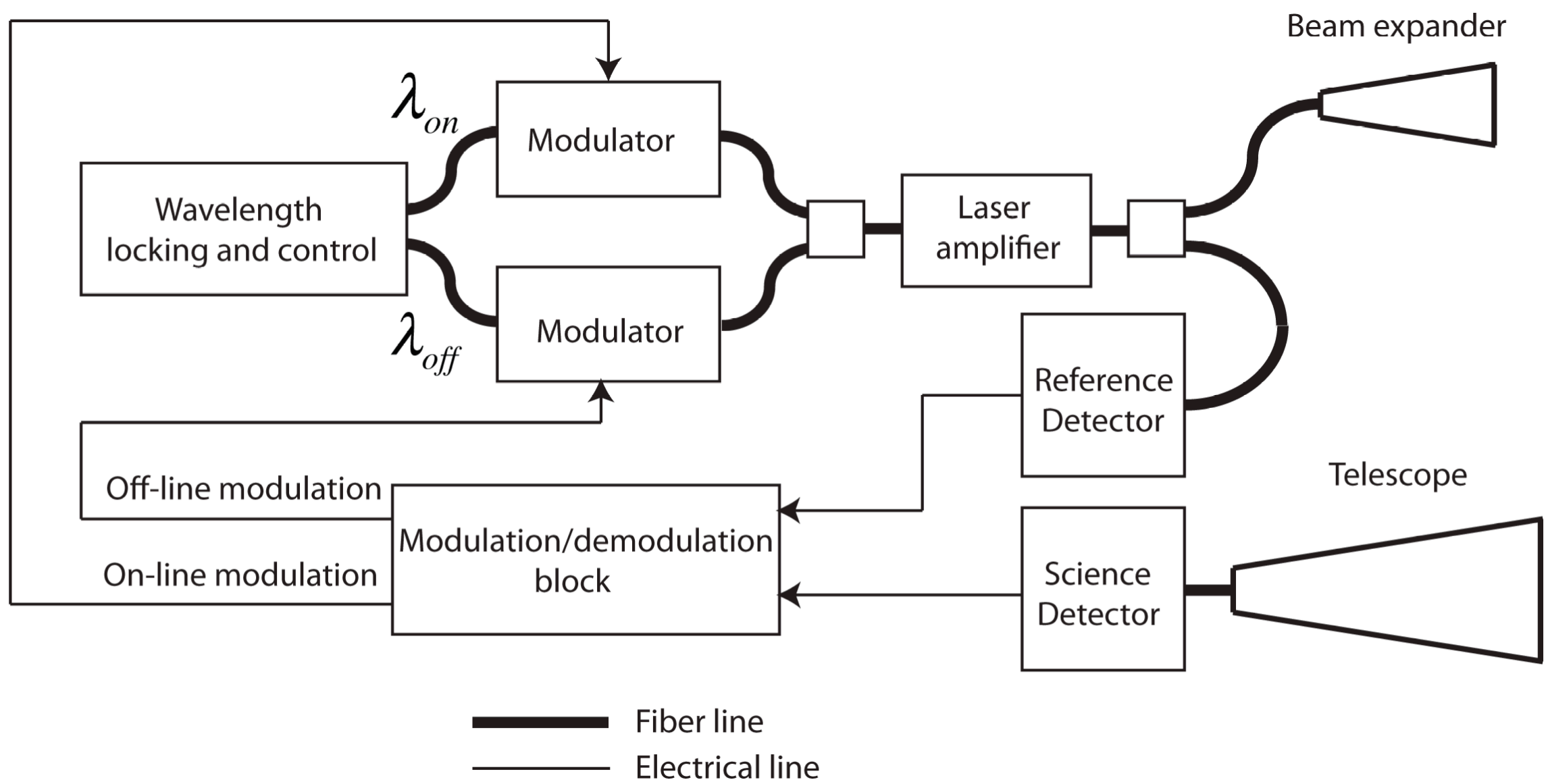

Figure 1. Simplified instrument block diagram.

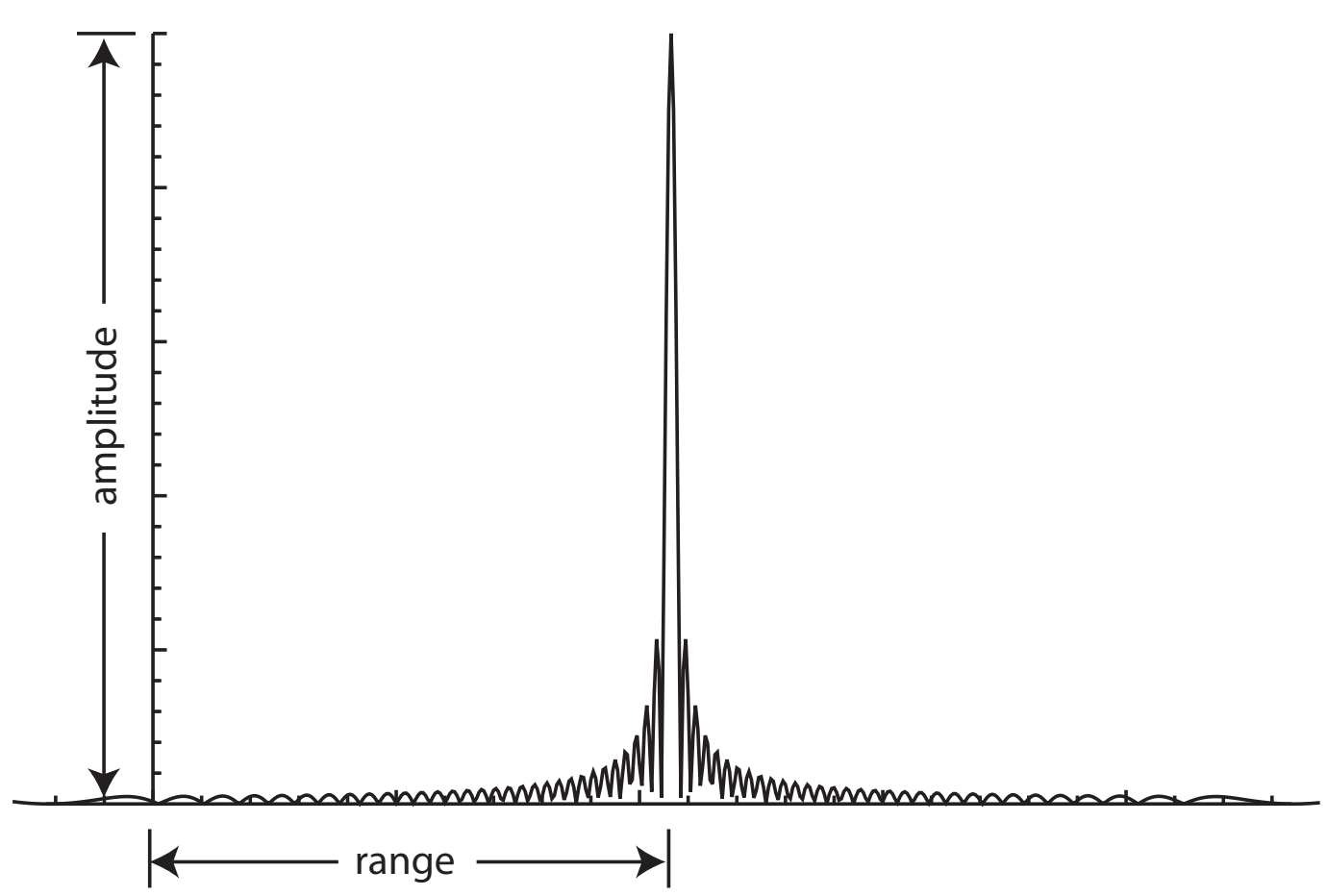

Figure 2. Result of cross correlation of the return signal to obtain range and amplitude.

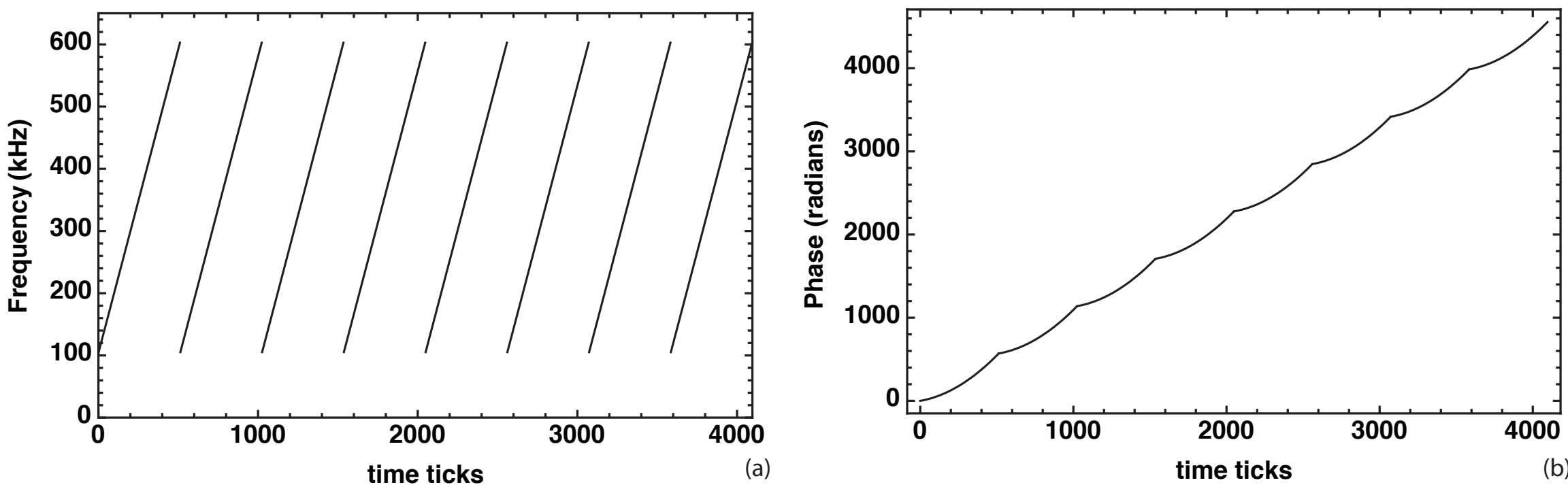

Figure 3. Frequency (a) and phase (b) as a function of time for linear sweep case.

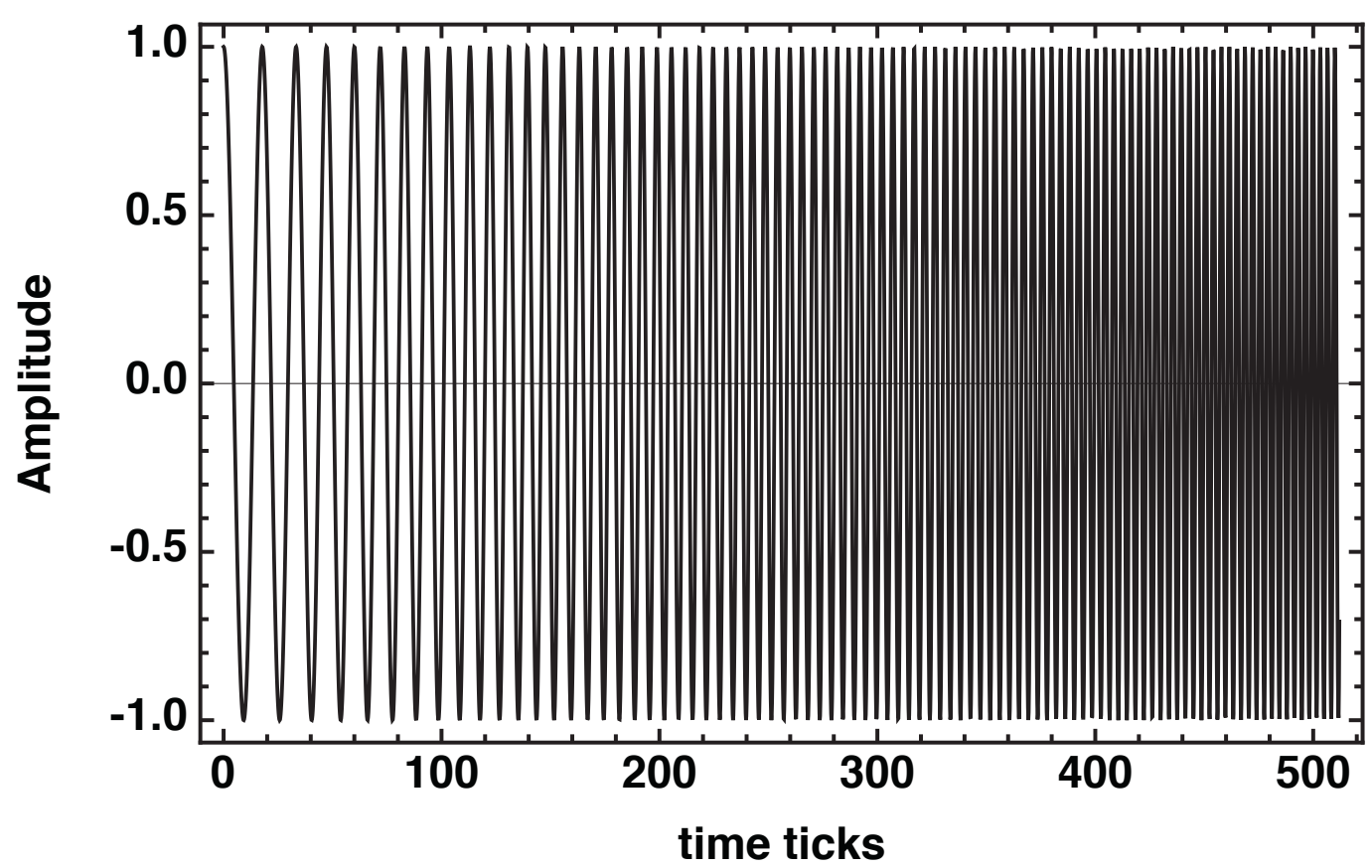

Figure 4. Single linear sweep sine wave.

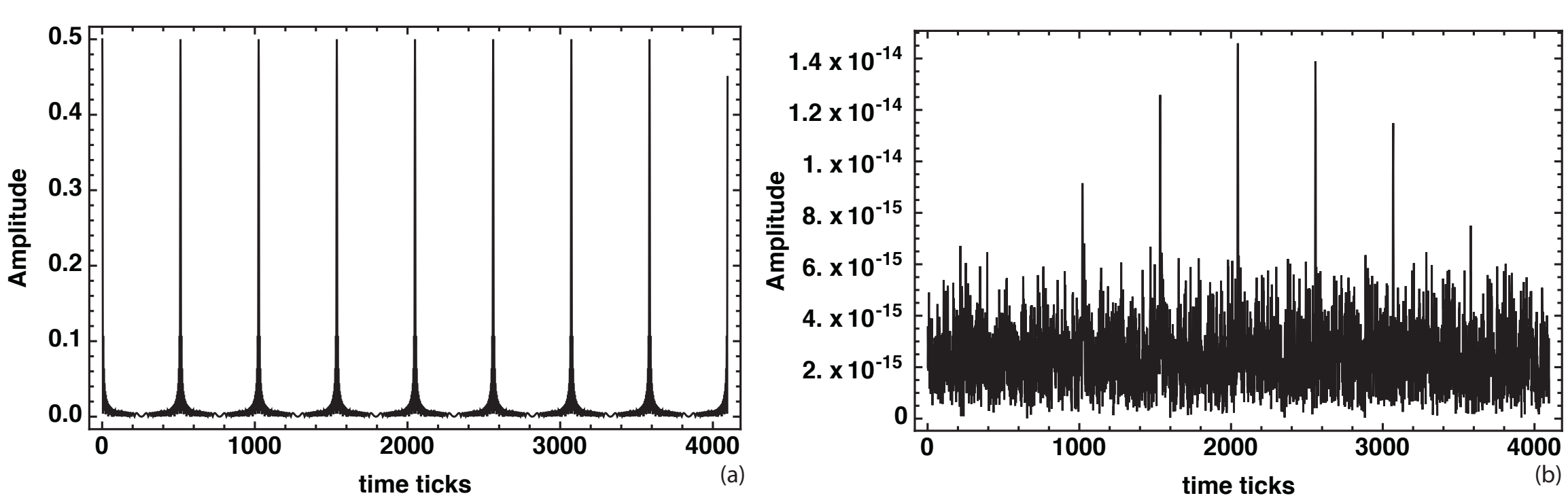

Figure 5. Channel 1 autocorrelation function (a) and channel 1-2 cross correlation cross talk (b).

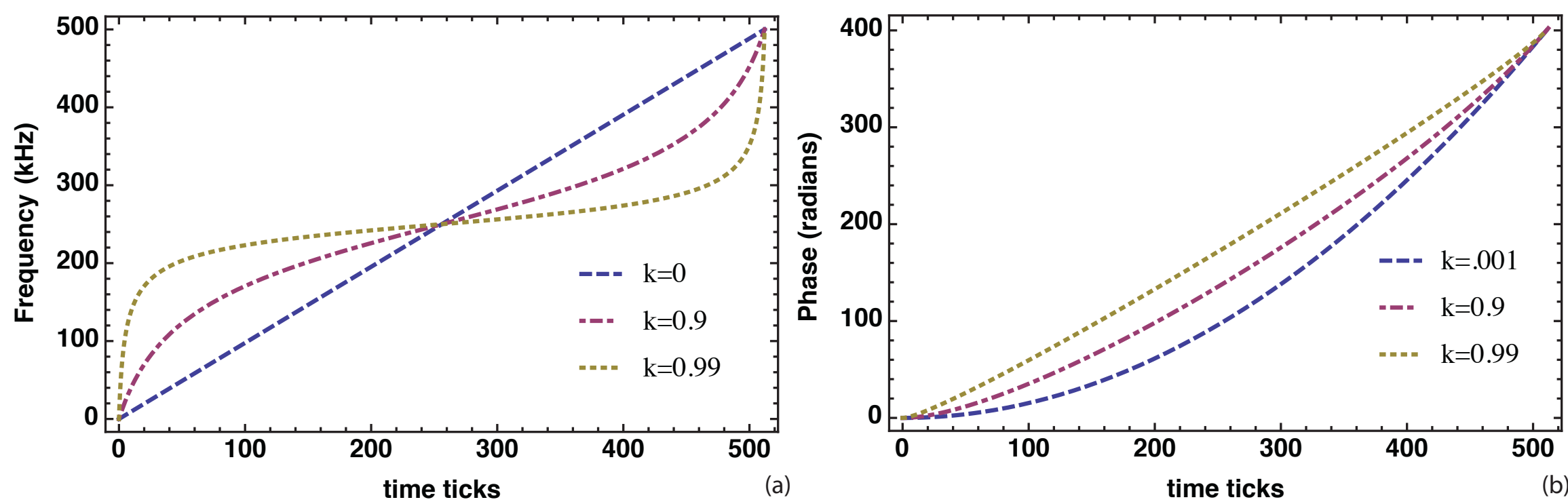

Figure 6. Single sweep nonlinear sweep (a) and phase (b) for different values of k.

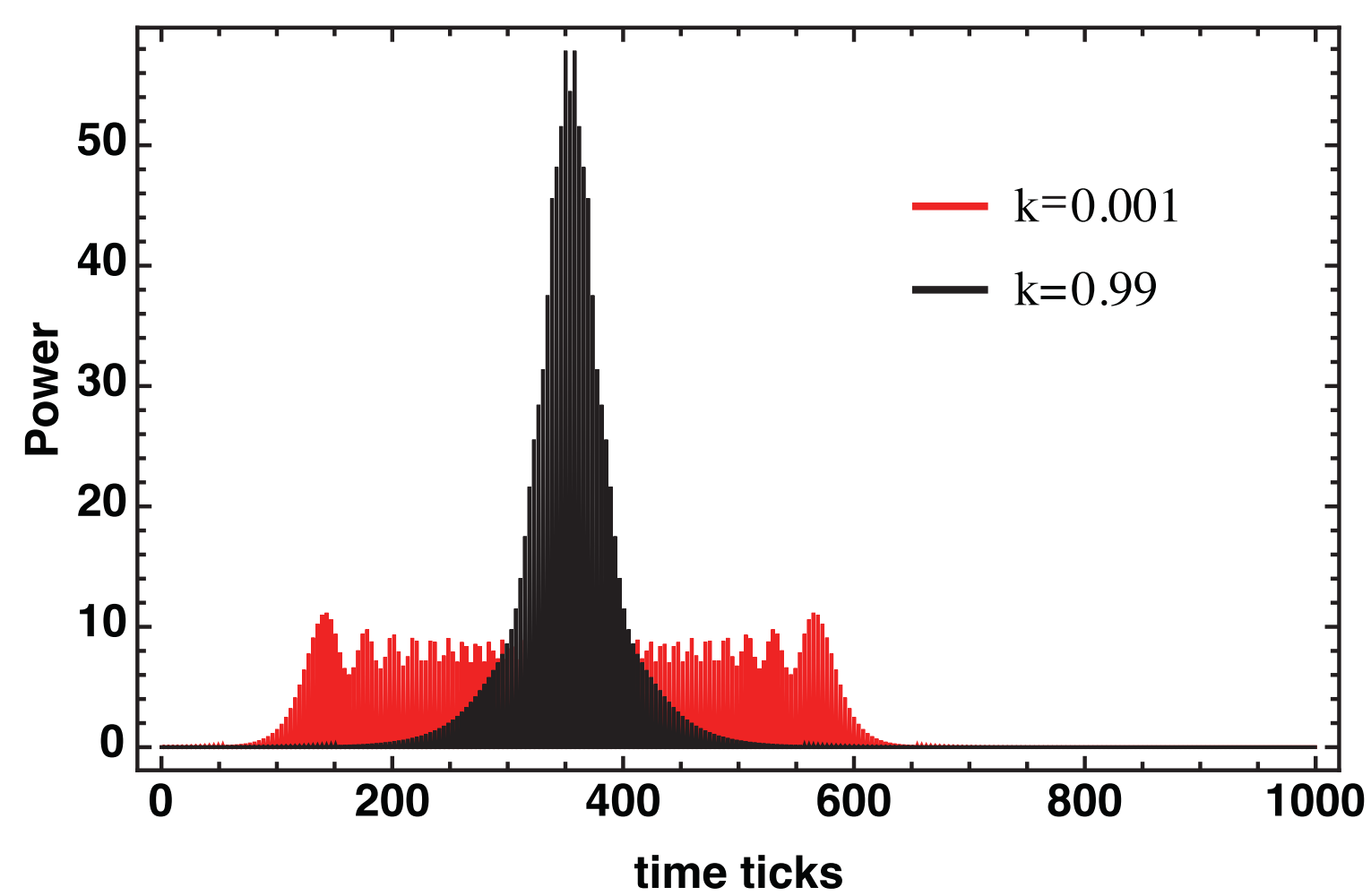

Figure 7. Comparison between linear and strongly non-linear case shows that as k approaches 1 the frequency profile approaches a Gaussian shape in a way that the area under the power curve is preserved.

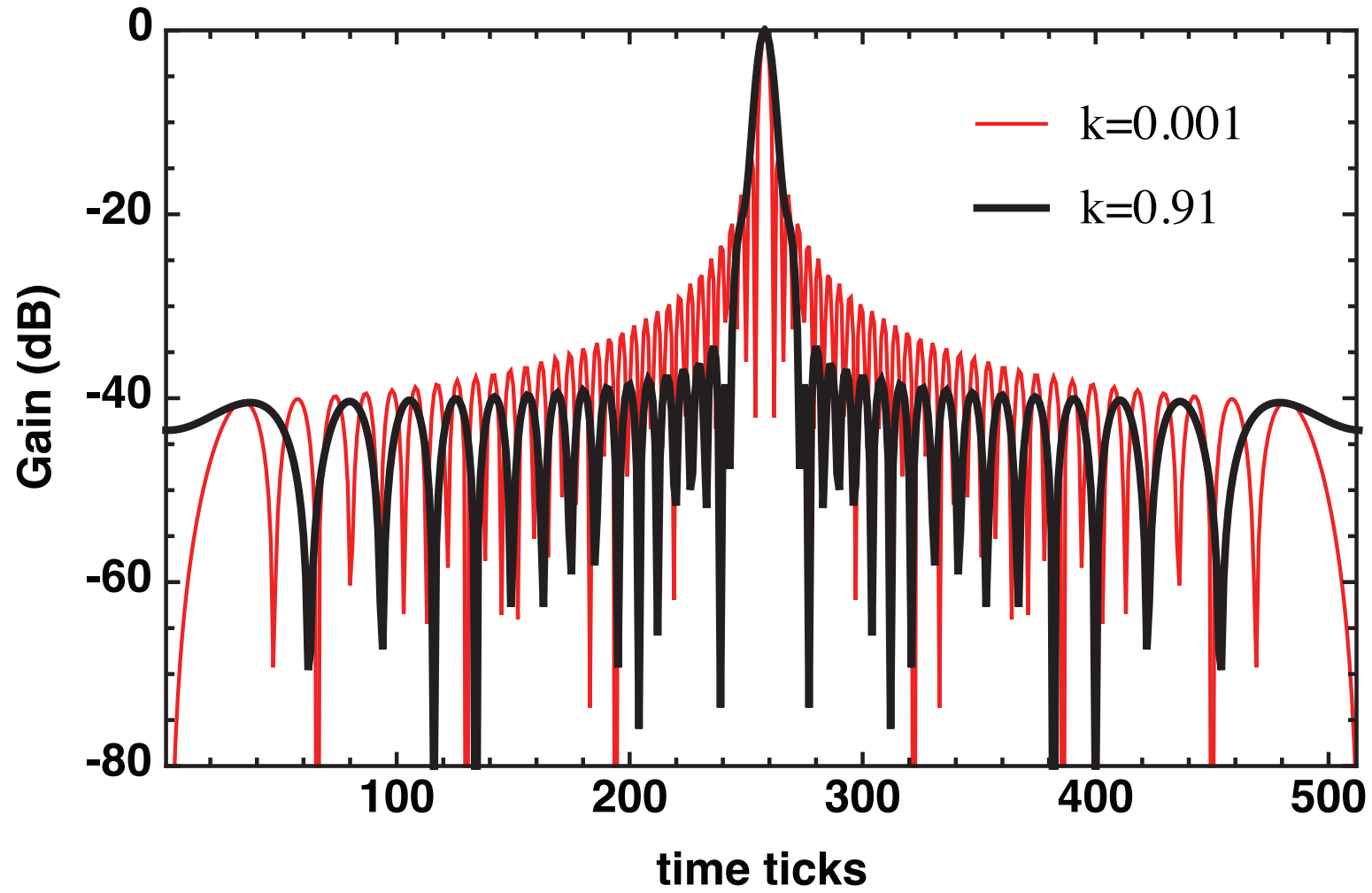

Figure 8. Comparison between linear sweep and non-linear sweep for 512 sweep length and k=0.91 shows highest sidelobe in non-linear case is about -34 dB compared to -13 dB in the linear case.

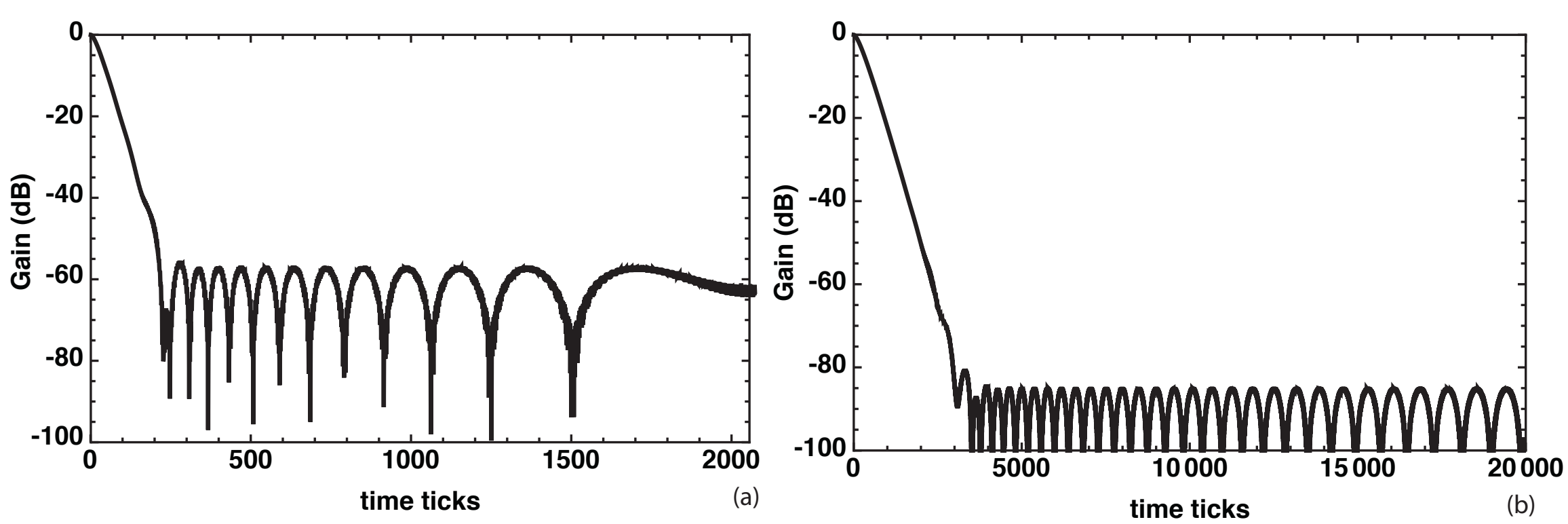

Figure 9. Highest sidelobe is -55 dB for k=0.999 and 4096 points per sweep (a). Highest sidelobe is below -80 dB for k=.99999 and $2^{17}$ points per sweep (b).

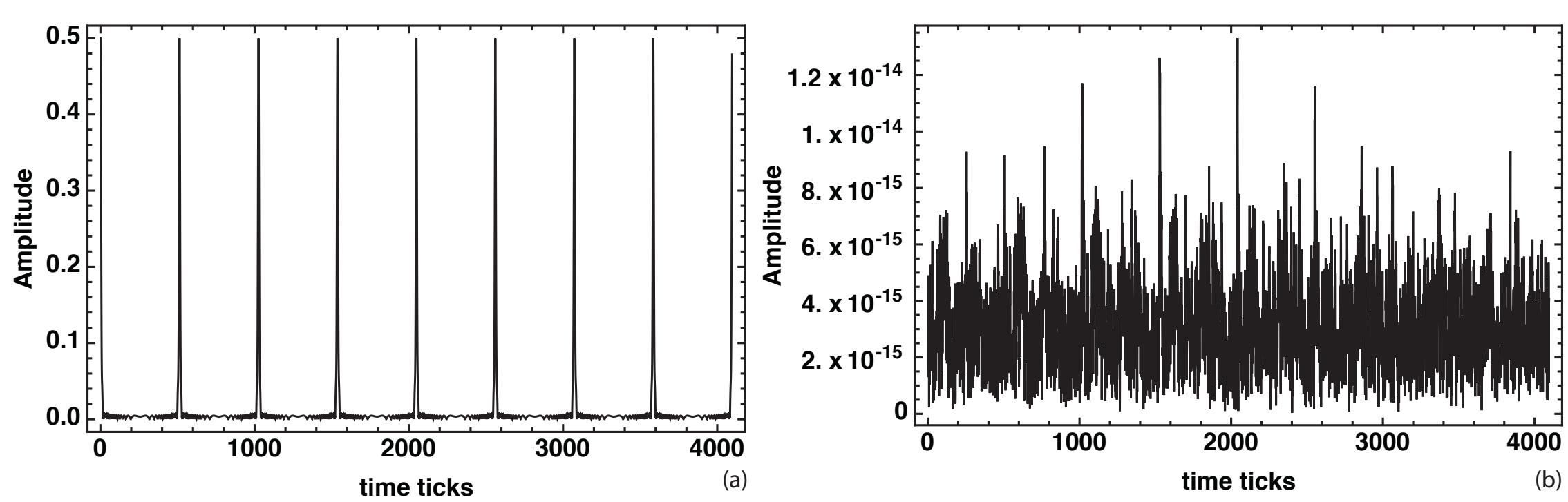

Figure 10. Channel 1 nonlinear autocorrelation function for k=0.91 and a sweep length of 512 shows reduced sidelobes (a) and channel 1-2 cross correlation cross talk is as good as the linear case (b).

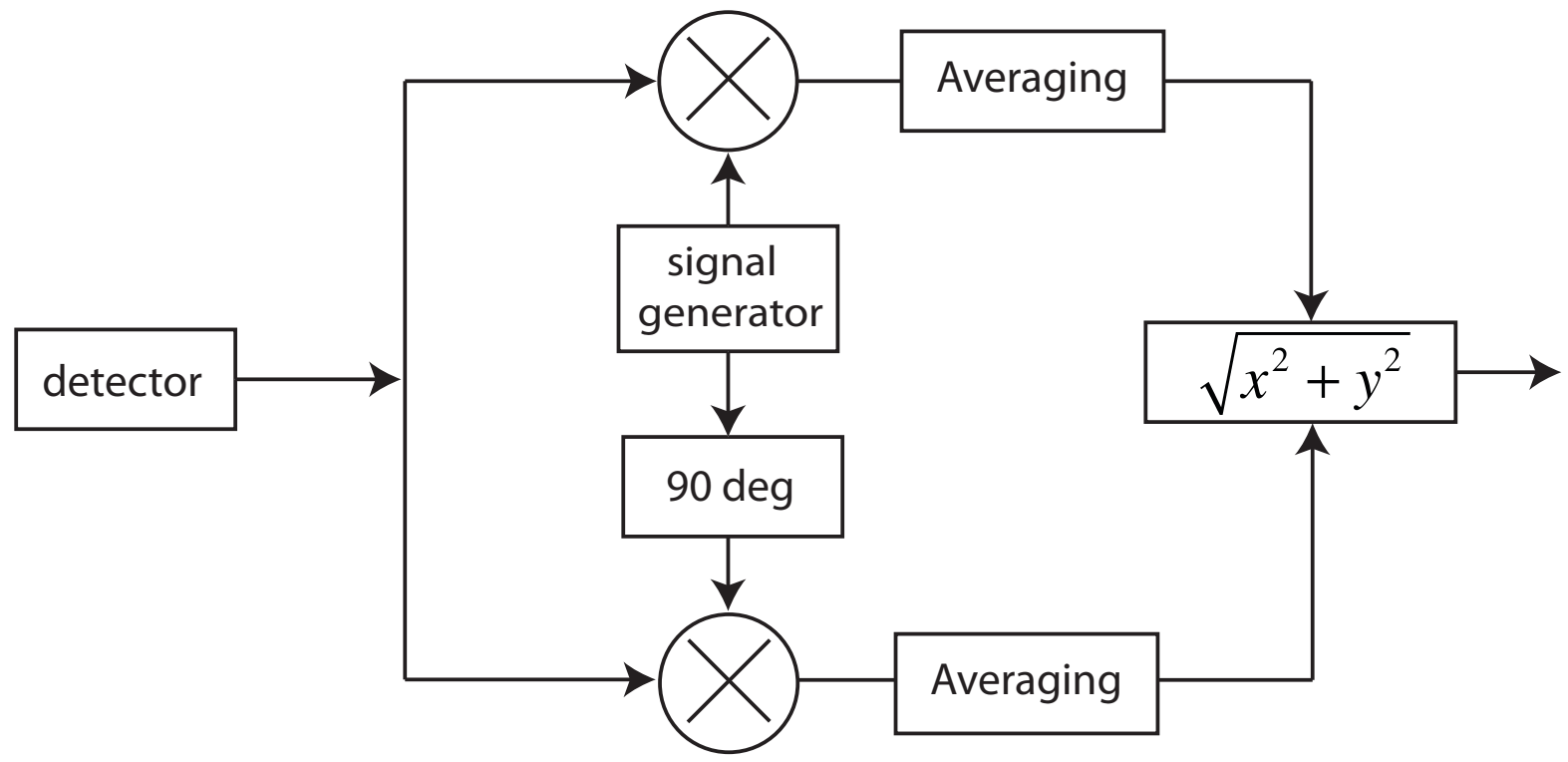

Figure A1. Quadrature/lock-in demodulation used for signal detection of unswept sine wave modulation.